\documentclass[journal=jctcce,manuscript=article]{achemso}

\usepackage{amsmath,amsfonts,amssymb}
\usepackage{graphicx}
\usepackage{setspace}
\usepackage{tocloft}
\usepackage{orcidlink}
\usepackage{booktabs}
\usepackage{lscape}
\usepackage[]{braket}
\usepackage{simplewick}
\usepackage{longtable}
\usepackage{rotating}
\usepackage{pdflscape}  

\usepackage{makecell}
\usepackage{xcolor}

\usepackage{amsmath,amssymb,bm}
\usepackage{graphicx}
\usepackage{booktabs}
\usepackage{multirow}
\usepackage{array}
\usepackage{siunitx}
\usepackage{float}
\usepackage{caption}
\usepackage{subcaption}
\usepackage{algorithm}
\usepackage{algpseudocode}
\usepackage{hyperref}
\usepackage{xcolor}
\usepackage{subcaption}
\usepackage[numbers,sort&compress]{natbib}
\usepackage{multicol}
\hypersetup{
    colorlinks=true,
    linkcolor=blue,
    citecolor=blue,
    urlcolor=blue
}

\usepackage[version=3]{mhchem} 

\author{Shamik Chanda\textsuperscript{\ddag}}
\author{Suhita Basumallick\textsuperscript{\ddag}}
\author{Sangita Sen}
\email{sangita.sen@iiserkol.ac.in}
\affiliation[IISER Kolkata]
{Department of Chemical Sciences and Centre for Advanced Functional Materials (CAFM), Indian Institute of Science Education and Research Kolkata, Mohanpur, Nadia-741246, West Bengal, India}

\title[An \textsf{achemso} demo]
  {Stall-Triggered Newton-Krylov-Assisted DIIS and Its Machine-Learning-Accelerated Variant for CCSD Convergence}

\abbreviations{IR,NMR,UV}

\begin{document}
\begingroup
\renewcommand{\thefootnote}{\ddag}
\footnotetext{These authors contributed equally to this work.}
\endgroup


\begin{abstract}
The iterative solution of the non-linear amplitude equations of coupled-cluster singles doubles (CCSD) theory can encounter convergence instabilities, asymptotic convergence or outright divergence. Direct Inversion in the Iterative Subspace (DIIS) is a widely used extrapolation technique but can also exhibit slow or oscillatory convergence in challenging regimes. In this work, we introduce a robust hybrid solver: Stall-Triggered Newton-Krylov Assisted DIIS (${}{\mathrm{st}}$NK-DIIS) alongside its machine-learning accelerated variant ${}{\mathrm{st}}$NK$^{\mathrm{ML}}$-DIIS which leverages the robust convergence of Newton-Krylov (NK) while avoiding the high computational cost of Krylov iterations. In our methods, DIIS remains the default accelerator and a matrix-free NK procedure is activated when sustained stagnation or divergence is detected with each NK step verified using the non-linear residual before acceptance. To alleviate the cost of Krylov iterations the ${}{\mathrm{st}}$NK$^{\mathrm{ML}}$-DIIS uses a gated recurrent unit (GRU) to predict a scalar logarithmic step-size ratio along the approximately computed Newton direction. When the ML-predicted step fails residual-based verification, the method falls back to a grid-based NK line search. Using CCSD as the underlying electronic-structure theory, we evaluate these methods against conventional DIIS and pure NK for stretched molecules exhibiting convergence difficulties in terms of residual convergence, iteration counts, and wall-time performance. These protocols may be generalizable to other non-linear equation solvers as well, as the ML model learns the convergence pattern of the variables and does not require system-specific knowledge such as atom identities or molecular orbital properties. This allows transferability across atom-type, bonding character and basis-sets.
\end{abstract}

\section{Introduction}
\label{intro}
In electronic structure theory, direct inversion in the iterative subspace (DIIS) provides an efficient history-based strategy for accelerating the iterative solution of non-linear amplitude equations\cite{Pulay1980,Pulay1982} such as coupled-cluster singles doubles (CCSD). By constructing an extrapolated solution from previously generated iterates and their associated residuals, DIIS exploits information accumulated along the convergence trajectory to reduce the number of iterations required for convergence while introducing relatively little computational overhead. However, in challenging convergence regimes, the information contained in the recent residual history may become insufficiently reliable to yield an effective extrapolation. Under such conditions, DIIS can generate overly aggressive or poorly directed updates, leading to oscillatory behavior, residual growth, or prolonged stagnation\cite{RohwedderSchneider2011,WalkerNi2011,EttenhuberJorgensen2015}. These limitations motivate the incorporation of an alternative correction mechanism that responds more directly to the local structure of the non-linear residual.\\

\noindent Newton–Krylov (NK) methods provide a complementary approach by constructing Newton corrections from the Jacobian of the non-linear residual equations and solving the resulting linear system iteratively using a Krylov method such as generalized minimal residual (GMRES)\cite{SaadSchultz1986,KnollKeyes2004,Yang2020NK}. By incorporating local Jacobian information, NK provides a problem-specific correction that is less dependent on the quality of the accumulated DIIS history. However, repeated evaluations of Jacobian–vector products, Krylov iterations, and non-linear residuals for step selection can make a pure NK strategy computationally demanding, particularly when applied throughout the entire convergence trajectory. These complementary characteristics motivate a hybrid strategy in which DIIS is retained as the default low-cost accelerator during normal convergence, while NK is invoked selectively when sustained stagnation or ineffective progress is detected to obtain a locally informed correction direction. We call this a stall-triggered ${}{\mathrm{st}}$NK-DIIS strategy.\\

\noindent Machine-learning (ML) approaches have increasingly been explored for molecular and electronic-structure problems, including the learning of potential-energy surfaces, molecular representations, and electronic properties \cite{BehlerParrinello2007,Rupp2012,Schutt2017,Welborn2018,Chmiela2017,Dral2020,Unke2021}. More recently, data-driven approaches have been investigated to accelerate coupled-cluster calculations and improve the efficiency of iterative CCSD procedures\cite{Margraf2018, Peyton2020,Townsend2019,TownsendVogiatzis2020,Ruth2022,Varuna2024,Agarawal2021,Allen2026, Pathirage2026}.  For example, Margraf and Reuter showed that vector representations constructed from cluster amplitudes can accurately learn coupled-cluster correlation energies, including from approximate perturbation-theory wavefunctions.\cite{Margraf2018} Extending this strategy, Peyton \textit{et al.} combined amplitude-based descriptors with a reduced density-matrix tensor representation to predict both coupled-cluster energies and one-electron properties such as molecular dipole moments.\cite{Peyton2020}. Townsend and Vogiatzis demonstrated that a machine learned starting guess as opposed to an MP2 can accelerate iterative CCSD convergence \cite{Townsend2019}. In subsequent work, they developed a transferable MP2-based machine-learning framework for predicting coupled-cluster energies\cite{TownsendVogiatzis2020}. Complementary approaches have explored data-driven strategies for selecting and sampling the two-electron excitation space, thereby reducing the training-data requirements and improving the applicability of ML-based coupled-cluster methods to larger systems and basis-sets\cite{Varuna2024}. Most recently, Pathirage and Vogiatzis introduced DDCCNet, a family of physics-enhanced multitask neural networks that simultaneously predict CCSD single- and double-excitation amplitudes and correlation energies from lower-level Hartree-Fock and MP2 features.\cite{Pathirage2026}. Agarawal et al. combined non-linear dynamical analysis with supervised machine learning to identify dominant cluster amplitudes and predict the remaining amplitudes, offering an alternative strategy for accelerating coupled-cluster calculations\cite{Agarawal2021}. Ruth et al. employed $\Delta$-learning to predict coupled-cluster perturbative triples-energy corrections, demonstrating another application of machine learning in correlated electronic-structure theory\cite{Ruth2022}. Further developments have extended coupled-cluster calculations to the generation of high-quality training data for machine-learning interatomic potentials\cite{Allen2026}. These studies establish a promising foundation for integrating machine learning into correlated electronic-structure calculations. The existing approaches focus on predicting cluster amplitudes, molecular properties, or final energies or reducing the cost of correlated calculations by reducing the number of variables to be iterated. In this work we propose to directly control the convergence behavior of the non-linear CCSD amplitude equations using solver-integrated machine-learning. Building on the stall-triggered NK-DIIS recovery framework (${}{\mathrm{st}}$NK-DIIS), we introduce a machine-learning extension, denoted ${}{\mathrm{st}}$NK$^{\mathrm{ML}}$-DIIS, in which  the ML model predicts only a scalar step-size quantity along the explicitly computed approximate Newton direction. A gated recurrent unit (GRU) processes the recent convergence history and predicts a logarithmic step-size ratio, reducing the machine-learning output from a high-dimensional amplitude-space prediction to a single scalar quantity. The predicted ratio is then used to construct an ML-guided NK update, which is evaluated using the non-linear residual before acceptance. Failing to reduce the residual as per the set criteria leads to the ML proposal being rejected, following which the computed Newton direction is retained, and a conventional grid-based NK line search is performed to determine a verified fallback step. Thus, the NK procedure provides the update direction, while the ML model determines the step magnitude along that direction, and the non-linear residual serves as the final acceptance criterion. Our developments motivate the use of machine learning not as a replacement for the underlying many-body equations, but as a mechanism for learning selected aspects of the iterative solution process.\\

\noindent The proposed framework is implemented using CCSD as the underlying electronic-structure theory and computational model. CCSD provides a well-established non-linear system of coupled amplitude equations\cite{Cizek1966,PurvisBartlett1982,Bartlett1989,BartlettMusial2007}and is a standard coupled-cluster method used in many applications in quantum chemistry \cite{Stanton1991,Raghavachari1989,CrawfordSchaefer2000,Scuseria1986}. It therefore provides a suitable and worthwhile framework for developing and evaluating iterative-solver strategies. We also emphasize that the methods discussed here address the convergence difficulties of the CCSD equations, not its accuracy or computational cost. In fact, we believe the methods may be amenable to other non-linear equations with training of the ML-model with suitable data as the method relies on detection of convergence patterns rather than electronic structure information.\\

\noindent  The ML model is trained on a set of small molecules at various geometries which provide challenging convergence regimes and tested on stretched H$_2$O (geometry not present in the training set), CO (geometry or basis set excluded from training set), CO$_2$, H$_2$CO, CH$_2$NH, BeH$_2$, and N$_2$O (molecules not present in training set) which also suffer from convergence issues. The training is economical with respect to memory and time as the training input is stall-focused. Only five iteration windows of each trajectory ending in a stall contribute to the supervised regression test. Moreover, only 19 extracted scalar features of the amplitude vectors need to be stored rather than the full amplitude space arrays. In contrast, Ref.\citenum{Townsend2019} used 30 electronic features per double excitation amplitude to enable amplitude prediction leading to growth in the training data with basis set improvement and number of electrons in the systems; our training data size is agnostic to the number of electrons in the systems and choice of basis set. However, the cost of the running the CCSD trajectories to generate the training data will scale as usual. Moreover, our method accelerates but does not bypass CCSD iterations which need to be computationally feasible. One may imagine that our method could help to generate training data corresponding to challenging convergence situations in small molecules and/or basis sets, as appropriate, which can then be fed into $\Delta$-learning techniques which allow prediction of CCSD energy when the system is too large for direct computation. The remainder of the paper is organized as follows. We first present the theoretical framework and working equations, followed by the computational implementation of the proposed methods. We then evaluate the convergence and computational performance of ${}{\mathrm{st}}$NK-DIIS and ${}{\mathrm{st}}$NK$^{\mathrm{ML}}$-DIIS in comparison with conventional DIIS and pure NK. Finally, we summarize the main findings of the study.

\section{Theoretical Formulation}
\label{theory}
\subsection{CCSD non-linear equations}
The CCSD wavefunction is represented as
\begin{equation}
|\Psi_{\mathrm{CCSD}}\rangle=e^{\hat{T}_1+\hat{T}_2}|\Phi_0\rangle
\end{equation}
where $\hat{T}_1$ and $\hat{T}_2$ are the single- and double-excitation operators. The corresponding non-linear amplitude equations can be written as,
\begin{equation}
\mathbf R(\mathbf T)=0
\end{equation}
where the single packed amplitude vector is $\mathbf T=
\begin{bmatrix}\mathbf T_1\\ \mathbf T_2\end{bmatrix}$ and the residuals can be written in the form
\begin{equation}
\mathbf{R}(\mathbf{T})=\mathbf{F}(\mathbf{T})-\mathbf{DT} \label{eq:ccsd}
\end{equation}
where $\mathbf{F}(\mathbf{T})$ 
denotes the non-linear right-hand side and $\mathbf{D}$ represents the diagonal denominator operator composed of the orbital-energy difference for the single and double excitation operators. The convergence measurement used throughout this work is the Euclidean norm of the residual,
\begin{equation}
\mathcal{R}=\|\mathbf{R}(\mathbf{T})\|_2    
\end{equation}
and convergence is declared when
\begin{equation}
\mathcal{R}<\mathcal{R}_{\mathrm{conv}}; \qquad \mathcal{R}_{\mathrm{conv}}=10^{-7}
\end{equation}
The CCSD equations remain unchanged by the proposed convergence-acceleration framework; the different strategies only modify the iterative procedure used to update the cluster amplitudes.

\subsection{Baseline DIIS iteration}
DIIS is retained as the default convergence-acceleration mechanism. From Eq. \ref{eq:ccsd}, the diagonal Jacobi state is obtained as
\begin{equation}
\mathbf{T}_{\mathrm{J}}=\mathbf{D}^{-1}\mathbf{F}(\mathbf{T})
\end{equation}
giving the Jacobi correction
\begin{equation}
\Delta \mathbf{T}_{\mathrm{J}}=\mathbf{T}_{\mathrm{J}}-\mathbf{T}  \label{eq:jacobi}
\end{equation}
Its Euclidean norm,
\begin{equation} \label{sJdef}
s_J=\|\Delta \mathbf{T}_{\mathrm{J}}\|_2    
\end{equation}
provides a local scale for subsequent Newton-step selection.\\

\noindent For a subspace containing $m$ previous amplitude states $\mathbf{T}_i$ and corresponding DIIS error vectors, $\{\mathbf{e}_i\}$, the DIIS coefficients are obtained from the equation
\begin{equation}
\begin{bmatrix}\mathbf{B} & -\mathbf{1}\\-\mathbf{1}^{T} & 0\end{bmatrix}\begin{bmatrix}\mathbf{c}\\\lambda\end{bmatrix}=\begin{bmatrix}\mathbf{0}\\-1\end{bmatrix}
\end{equation}
where
\begin{equation}
B_{ij}=\mathbf{e}_i^T \mathbf{e}_j    
\end{equation}
The extrapolated DIIS amplitude vector is then
\begin{equation}
\mathbf{T}_{\mathrm{DIIS}}=\sum_i c_i\mathbf{T}_i
\end{equation}
The associated DIIS correction used in the feature representation for the ML model is defined as 
\begin{equation} \label{DeltaTDIIS}
 \Delta \mathbf{T}_{\mathrm{DIIS}}=\mathbf{T}_{\mathrm{DIIS}}-\mathbf{T}   
\end{equation}
Thus, using information accumulated from previous iterations, DIIS provides the primary low-cost acceleration mechanism during the normal convergence trajectory. The DIIS history is stored consistently throughout the calculation and is also retained after a successful NK recovery.

\subsection{Newton-Krylov recovery}
For the non-linear Eq. \ref{eq:ccsd}, the Newton correction ($\Delta \mathbf{T}_{\mathrm{N}}$) satisfies 
\begin{equation}
\mathbf{J}(\mathbf{T})\Delta \mathbf{T}_N=-\mathbf{R}(\mathbf{T}) ; \quad \mathbf{J}(\mathbf{T})=\frac{\partial \mathbf{R}}{\partial \mathbf{T}}
\end{equation}
where $\mathbf{J}(\mathbf{T})$ is the Jacobian of the non-linear residual.\\

\noindent Rather than explicitly constructing the Jacobian, the present implementation evaluates Jacobian-vector products in matrix-free form as,
\begin{equation}
\mathbf{J}(\mathbf{T})\mathbf{v}\approx \frac{\mathbf{R}(\mathbf{T}+h\mathbf{v})-\mathbf{R}(\mathbf{T})}{h}.
\end{equation}
where, h is a finite-difference displacement. GMRES is then used to iteratively obtain an approximate Newton correction, $\Delta \mathbf{T}_{\mathrm{N}}$. \\

\noindent At the beginning of each iteration, the nonlinear residual is evaluated to determine the current convergence state. This same residual is passed through the NK procedure and reused in the matrix-free GMRES calculation, avoiding an unnecessary duplicate evaluation of the starting residual. As the present framework, separates the Newton direction from its magnitude, the resulting correction is normalized according to
\begin{equation}
\mathbf{v}_N=\frac{\Delta \mathbf{T}_N}{\|\Delta \mathbf{T}_N\|_2}; \qquad \|\mathbf{v}_N\|_2=1 
\end{equation}
to obtain the Newton direction \textbf{v}$_N$. Consequently, NK determines the locally informed correction direction, while the magnitude of the step is selected separately. 

\subsection{Stall-triggered NK-DIIS}
To avoid the computational cost of applying NK at every iteration, the method monitors the convergence behavior of the accepted residual trajectory. The residual ratio between two successive iterations is defined as
\begin{equation}
r_k=\frac{\mathcal{R}_{k+1}}{\mathcal{R}_k}
\end{equation}
A stall is detected when
\begin{equation}
 r_k >\rho_{\mathrm{stall}} 
\end{equation}
for $N_{\mathrm{stall}}$ consecutive iterations. The parameters used in the present implementation are
\begin{equation*}
N_{\mathrm{stall}}=3, \qquad \rho_{\mathrm{stall}}=0.98.
\end{equation*}
Thus, NK recovery is activated only after sustained residual stagnation rather than on every iteration.\\

\noindent When a stall is detected, a Newton direction, $\mathbf{v}_N$, is computed from the current state. Let $\mathbf T_{\mathrm{before}}$ denote the amplitude vector frozen immediately before the NK trial. The used step-sizes s$_j$ are constructed relative to the norm of the Jacobi correction s$_J$ defined in Eq.~\ref{sJdef},
\begin{equation}
s_j=f_js_J; \quad f_j\in \{0.125,0.25,0.5,1.0,1.5,2.0\}
\end{equation}
The corresponding amplitude states are represented as
\begin{equation}
\mathbf{T}_j=\mathbf{T}_{\mathrm{before}}+s_j\mathbf{v}_N
\end{equation}
For each step-size, the residual norm is evaluated,
\begin{equation}
 \mathcal{R}_{j}=\|\mathbf{R}(\mathbf{T}_{j})\|_2   
\end{equation}
The step-size producing the smallest finite residual on the prescribed grid is identified as,
\begin{equation}
s_{\mathrm{NK}}=\underset{s_j}{\operatorname{argmin}}\;\mathcal{R}_j
\end{equation},
giving,
\begin{equation}
\mathbf{T}_{NK}=\mathbf{T}_{\mathrm{before}}+s_{NK}\mathbf{v}_N
\end{equation}
and,
\begin{equation}
 \mathcal{R}_{NK}=\|\mathbf{R}(\mathbf{T}_{NK})\|_2.
\end{equation}
The selected Newton state is accepted only if it provides a genuine residual reduction relative to the frozen pre-recovery state. In the present implementation, the required relative reduction is
\begin{equation}
\frac{\mathcal{R}_{\mathrm{before}}-\mathcal{R}_{\mathrm{NK}}}{\mathcal{R}_{\mathrm{before}}}\geq \eta_{\mathrm{NK}}, \qquad \eta_{\mathrm{NK}}=10^{-3}   
\end{equation}
together with
\begin{equation}
\mathcal{R}_{\mathrm{NK}}<\mathcal{R}_{\mathrm{before}}
\end{equation}
If the acceptance conditions are not satisfied, the entire NK trial is discarded and the amplitudes are restored to $\mathbf T_{\mathrm{before}}$. The solver then proceeds through the ordinary Jacobi/DIIS pathway.\\

\noindent Following a successful NK recovery, the DIIS history is retained rather than discarded, allowing the history-based accelerator to continue from the recovered trajectory. To prevent repeated NK correction and in order to return control to DIIS, the implementation also forces a DIIS recalibration step after a prescribed number of consecutive successful NK recoveries. The present default is $N_{\mathrm{verify}}=3$. After this number of successful NK steps, the next iteration is forced through the DIIS pathway, after which normal stall monitoring resumes.\\

\noindent The resulting strategy therefore invokes the more expensive NK calculation only when the inexpensive DIIS procedure exhibits sustained stagnation, while allowing successful Newton corrections to return the solver to the normal DIIS trajectory.

\subsection{ML-assisted Newton step-size selection}
The ML extension does not replace the NK correction direction \textbf{v}$_N$. Instead, it predicts the analogue of s$_j$, denoted as s$_{ML}$. However, it does not do so directly. 
Rather, the oracle predicts a scalar logarithmic step size ratio $q ^{pred}$,
\begin{equation}
q ^{pred}=\log_{10}\left(\frac{s_{ML}}{s_J}\right) \label{eq:ml}
\end{equation}
such that,
\begin{equation}
s_{\mathrm{ML}}=s_J10^{q^{pred}}
\end{equation}
and
\begin{equation}
\Delta \mathbf{T}_{\mathrm{ML}}=s_{\mathrm{ML}}\mathbf{v}_N
\end{equation}
Since $\mathbf{v}_N$ is normalized,
\begin{equation}
\|\Delta \mathbf{T}_{\mathrm{ML}}\|_2=s_{\mathrm{ML}}
\end{equation}
The ML model, therefore, has a single scalar output: the logarithmic magnitude of the Newton step relative to the local Jacobi-step scale. The high-dimensional direction in amplitude space remains explicitly determined by the NK calculation.\\

\noindent This separation is central to the proposed method: NK determines the correction direction, while ML determines how far to move along that direction.

\subsubsection{Feature representation and GRU-MLP model}
\label{feature}
At each iteration, the current CCSD state is characterized using 19 global diagnostics describing both the instantaneous state and recent convergence history. The feature vector is represented as

\begin{align*}
\mathbf{x}_k &= \Big[\log_{10}|R_k|, \mathrm{N}_{\mathrm{DIIS}}, \mathrm{std}(R_k/D), \max|R_k/D|, \mathrm{std}(T_k), \max|T_k|, \log_{10}|R_{k-1}|, \log_{10}|R_{k-2}|, \\
&\quad \log_{10}|R_{k-3}|, \log_{10}\left|\frac{R_k}{R_{k-1}}\right|, \log_{10}\left|\frac{R_{k-1}}{R_{k-2}}\right|, \log_{10}|E_{k-1}|, \log_{10}|E_{k-2}|, \log_{10}\left\|\Delta T_J\right\|_2, \\
&\quad \log_{10}\left\|\Delta T_{\mathrm{DIIS}}\right\|_2, \log_{10}\left\|\Delta T_{\mathrm{DIIS}}-\Delta T_J\right\|_2, \log_{10}\left\|\mathbf{e}_k\right\|_2, \log_{10}|D|_{\min}, \log_{10}|D|_{\max}\Big]
\end{align*}

Where, $R_k$ is the residual, $D$ is the denominator array, $D^{-1}R_k$ denotes the element-wise denominator-scaled residual, $N_{\mathrm{DIIS}}$ is the current DIIS subspace size, $\Delta T_{\mathrm{DIIS}}$ is the DIIS-predicted amplitude update (Eq.~\ref{DeltaTDIIS}) and $\mathbf{e}_k$ refers to the DIIS error vector associated with the current iteration. Absolute values are used for signed quantities, including the energy-related features and orbital-energy denominators.\\

\noindent To incorporate temporal information, a five-iteration window is constructed,
\begin{equation}
\mathbf{X}_k=[\mathbf{x}_{k-4},\mathbf{x}_{k-3},\mathbf{x}_{k-2},\mathbf{x}_{k-1},\mathbf{x}_k]; \quad \mathbf{X}_k\in \mathbb R^{5\times19}
\end{equation}

\noindent The ML architecture was chosen to match the temporal and low-dimensional nature of the step-selection problem. The five-step temporal window allows the model to account for the recent evolution of the convergence trajectory rather than relying only on the instantaneous solver state. A gated recurrent unit (GRU) is therefore used to encode the short-range temporal information, while a small multilayer perceptron (MLP) serves as the regression head for predicting the single scalar step-size parameter in Eq. \ref{eq:ml}.\\

\noindent  Several alternative architectures are also possible for the present sequence-to-scalar learning problem. A conventional MLP provides a direct mapping from the solver descriptors to the scalar target, but does not explicitly represent the temporal ordering of the recent convergence history. Recurrent neural network architectures such as long short-term memory (LSTM) networks provide an alternative means of encoding temporal information, although their additional gating mechanisms are not essential for the short five-iteration history which is considered here. One-dimensional convolutional networks can capture local patterns within the temporal sequence, while Transformer-based architectures can represent interactions across different time steps. However, these approaches introduce additional architectural complexity relative to the dimensionality and temporal extent of the present input. Graph neural networks (GNN) are primarily motivated by representations involving molecular connectivity or atom-level environments and are therefore less directly aligned with the present formulation, in which the input consists solely of global convergence diagnostics. A systematic comparison with these alternative architectures is beyond the scope of the present study. \\

\noindent The GRU–MLP mapping is given by
\begin{align*}
 &\mathbf{H}=\mathrm{GRU}(\mathbf{X}_k) \in \mathbb{R}^{32\times 5}, \qquad \mathbf{h}=\mathbf{H}_5 \in \mathbb{R}^{32\times 1}, \\ 
 & z=\mathbf{W}_2\sigma(\mathbf{W}_1\mathbf{h}+\mathbf{b}_1)+\mathbf{b}_2 \in \mathbb{R}   
\end{align*}
Where $\mathbf{H}=(\mathbf{H}_1,\dots,\mathbf{H}_5)$ are the hidden states of a single-layer
GRU of hidden dimension 32 after reading each of the five iterations, and $\mathbf{h}=\mathbf{H}_5$ is the final state, which summarizes the whole window. The MLP head has weights $\mathbf{W}_1\in\mathbb{R}^{32\times 32}$,
$\mathbf{W}_2\in\mathbb{R}^{1\times 32}$, biases $\mathbf{b}_1\in\mathbb{R}^{32}$,
$\mathbf{b}_2\in\mathbb{R}$, and SiLU activation $\sigma(x)=x/(1+e^{-x})$ acting in each element of $(\mathbf{W}_1\mathbf{h}+\mathbf{b}_1)$.
The predicted logarithmic step ratio is bounded using
\begin{equation}
q^{pred}=q_{\max}\tanh(z), \qquad q_{\max}=3
\end{equation}
which gives
\begin{equation}
-3\leq q^{pred}\leq3, \quad 10^{-3}s_J\leq s_{\mathrm{ML}}\leq10^3s_J   
\end{equation}
The bounded output restricts the predicted Newton step to a finite range, preventing arbitrarily large or excessively small corrections.

\subsubsection{Training and live-step verification}
\label{training}
Training data are generated from genuine stalled CCSD trajectories. For each stalled configuration, the oracle procedure evaluates the minimal residual along the normalized Newton direction,
\begin{equation}
s^*=\underset{s_j}{\operatorname{argmin}}\;\|\mathbf{R}(\mathbf{T}_{\mathrm{before}}+s_j\mathbf{v}_N)\|_2
\end{equation}
where the minimization is calculated over a predefined broad oracle step grid spanning the allowed step-ratio range, $-3\leq q^*\leq3$. The corresponding scalar training target is
\begin{equation}
q^*=\log_{10}\left(\frac{s^*}{s_J}\right).
\end{equation}
Thus, the training target represents the best step scale identified by the discrete oracle search rather than an analytically exact line-search minimum as in the conventional NK procedure.\\

\noindent The ML model is trained by minimizing the mean-squared error,
\begin{equation}
\mathcal{L}(\theta)=\frac{1}{N}\sum_{i=1}^{N}( q_i^{pred}-q_i^*)^2
\end{equation}
where $\theta$ denotes the trainable model parameters and $N$ represents the number of stalled trajectories used for training.\\

\noindent During live execution, the GRU supplies $q^{pred}$ and the predicted ML amplitude is constructed from the frozen pre-recovery state and $s_{ML}=s_{J} 10^{q^{pred}}$,
\begin{equation}
\mathbf{T}_{\mathrm{ML}}=\mathbf{T}_{\mathrm{before}}+s_{\mathrm{ML}}\mathbf{v}_N   
\end{equation}
Its residual norm $\mathcal{R}_{\mathrm{ML}}$ is then evaluated as,
\begin{equation}
\mathcal{R}_{\mathrm{ML}}=\|\mathbf{R}(\mathbf{T}_{\mathrm{ML}})\|_2.
\end{equation}
The ordinary DIIS residual R$_{DIIS}$ is also explicitly evaluated when an ML recovery is attempted, providing a direct baseline against which the learned proposal can be tested. The ML proposal is accepted only when it improves upon both the current state and the ordinary DIIS candidate,
\begin{equation}
\mathcal{R}_{\mathrm{ML}}<\mathcal{R}_{\mathrm{current}} \quad \& \quad \mathcal{R}_{\mathrm{ML}}<\mathcal{R}_{\mathrm{DIIS}}
\end{equation}
or equivalently,
\begin{equation}
 \mathcal{R}_{\mathrm{ML}}<\min\left(\mathcal{R}_{\mathrm{current}},\mathcal{R}_{\mathrm{DIIS}}\right).   
\end{equation}
This verification step ensures that the learned prediction cannot override the nonlinear residual criterion. Thus ML produces an improved step and is a true DIIS accelerator.\\

\noindent If the proposed ML is rejected, the same Newton direction, $\mathbf{v}_N$, is then reused in a grid-based fallback  line search using
\begin{equation}
s_j\in
\{0.25s_J,0.5s_J,1.0s_J,1.5s_J,2.0s_J\}
\end{equation}
The fallback step is selected as
\begin{equation}
s_{\mathrm{NK}}=\underset{s_j}{\operatorname{argmin}}
\;\|\mathbf{R}(\mathbf{T}+s_j\mathbf{v}_N)\|_2
\end{equation}
and is accepted only when
\begin{equation}
\mathcal{R}_{\mathrm{NK}}<\mathcal{R}_{\mathrm{current}}.
\end{equation}
For stNK$^{ML}$-DIIS, the present default is $N_{\mathrm{verify}}=2$, after which DIIS is restored after successful NK recovery.
If both the ML proposal and the fallback Newton step fail their residual-based acceptance tests, the solver restores, $\mathbf T_{\mathrm{before}}$, and proceeds through the ordinary Jacobi/DIIS pathway.\\

\noindent Importantly, the Newton direction is computed only once for a given ML recovery attempt and is shared by the ML proposal and the subsequent fallback search. Consequently, rejection of the ML step does not require a second GMRES calculation. The distinction between the two searches is intentional. The training procedure uses a broad oracle grid to expose the model to a wide range of useful step magnitudes, whereas the live fallback uses a smaller conservative grid to provide a low-cost recovery mechanism when the learned proposal is rejected.

\subsection{Overall computational workflow}

\noindent The complete procedure for ${}{\mathrm{st}}$NK$^{\mathrm{ML}}$-DIIS is summarized in Algorithm 1. The resulting framework separates the roles of acceleration, recovery, and step selection: DIIS provides the default acceleration, stall-triggered NK supplies the recovery direction, and the ML model  selects the recovery step magnitude. In all cases, the nonlinear CCSD residual remains the final acceptance criterion for a recovery update. The underlying CCSD nonlinear equations remain unchanged throughout. The proposed framework therefore modifies the iterative path used to solve the nonlinear amplitude equations, not the equations themselves.

\begin{algorithm}[H]
\caption{${}{\mathrm{st}}$NK$^{\mathrm{ML}}$-DIIS}
\label{alg:mlstnk}
\begin{algorithmic}[1]
\State Initialize CCSD amplitudes $T$
\State Initialize DIIS and feature histories
\State Load trained scalar-step predictor
\For{$k=1,\ldots,k_{\max}$}
\State Freeze current amplitudes: $T_{\mathrm{before}}\gets T$
\State Compute nonlinear RHS $F(T)$
\State Compute residual $R=F-DT$
\State $\mathcal{R}_{\mathrm{before}}\gets\|R\|_2$
\If{$\mathcal{R}_{\mathrm{before}}<\mathcal{R}_{\mathrm{conv}}$}
\State Declare convergence
\State \textbf{break}
\EndIf
\State Construct Jacobi state $T_J$
\State $\Delta T_J\gets T_J-T_{\mathrm{before}}$
\State $s_J\gets\|\Delta T_J\|_2$

\State Construct and store the current 19-dimensional feature vector
\State Append feature vector to the temporal window

\If{DIIS is stalled and feature window is full}
    \State Construct DIIS trial $T_{\mathrm{DIIS}}$
    \State Evaluate $\mathcal{R}_{\mathrm{DIIS}}$
    \State Restore $T_{\mathrm{before}}$
    \State Compute normalized Newton-Krylov direction $v_N$
    \State Predict $q^{pred}=\mathrm{ML}(X_{k-4:k})$
    \State $s_{\mathrm{ML}}\gets s_J10^{q^{pred}}$
    \State $T_{\mathrm{ML}}\gets T_{\mathrm{before}}+
    s_{\mathrm{ML}}v_N$
    \State Evaluate $\mathcal{R}_{\mathrm{ML}}$
    \If{$\mathcal{R}_{\mathrm{ML}}<
    \min(\mathcal{R}_{\mathrm{before}},
    \mathcal{R}_{\mathrm{DIIS}})$}
        \State Accept $T_{\mathrm{ML}}$
        \State Record ML acceptance
    \Else
        \State Search fallback trial scales along the same $v_N$
        \State Select the best finite fallback residual
        \If{$\mathcal{R}_{\mathrm{NK}}<
        \mathcal{R}_{\mathrm{before}}$
        \textbf{and}
        $\displaystyle
        \frac{\mathcal{R}_{\mathrm{before}}-\mathcal{R}_{\mathrm{NK}}}
        {\mathcal{R}_{\mathrm{before}}}
        \geq 10^{-3}$}
            \State Accept NK fallback
        \Else
            \State Restore $T_{\mathrm{before}}$
            \State Apply ordinary DIIS/Jacobi update
        \EndIf
    \EndIf
\Else
    \State Apply ordinary DIIS/Jacobi update
\EndIf

\State Evaluate post-update residual and energy change
\State Update convergence and stall histories
\State If required, force DIIS recalibration after two consecutive successful NK recoveries
\EndFor
\end{algorithmic}
\end{algorithm}

\section{Implementation}
\label{implementation}
All convergence algorithms developed in this work were implemented in an original, standalone Python solver suite\cite{Chanda2026stNKDIIS} interfaced with the Psi4 quantum-chemistry package\cite{Parrish2017} via its Python API (PsiAPI). Psi4 is used solely to obtain the reference Hartree-Fock wavefunction and molecular-orbital integrals, while only the spin-orbital CCSD right-hand-side amplitude equations in \texttt{helper\_CC.py} were adapted from the Psi4NumPy reference implementation\cite{Smith2018Psi4NumPy}. All iterative solver components--including the unified solver driver, conventional DIIS, matrix-free finite-difference Jacobian-vector products, diagonally preconditioned GMRES (via SciPy) for Newton-Krylov recovery, RHS-caching line search, stall-detection logic, $\mathrm{st}\text{NK-DIIS}$, and the PyTorch-based GRU training and inference pipeline for $\mathrm{st}\text{NK}^{\mathrm{ML}}\text{-DIIS}$--were developed originally for this study\cite{Chanda2026stNKDIIS}.

All four benchmarked convergence strategies (conventional DIIS, pure NK, $\mathrm{st}\text{NK-DIIS}$, and $\mathrm{st}\text{NK}^{\mathrm{ML}}\text{-DIIS}$) are executed through this common driver using identical integral preparation, convergence criteria ($R_{\mathrm{conv}}=10^{-7}$), and computational settings; only the amplitude-update procedure differs across methods. (While the archived driver also supports additional iterative variants and UHF references under the same interface, the present study focuses exclusively on these four closed-shell RHF schemes.) During live runs, the GRU predictor is evaluated on the CPU within the same Python execution environment as the CCSD solver. Thus, the reported differences in iteration counts and wall-time results directly reflect the convergence algorithms rather than differences in underlying integral or computational settings.

\noindent All calculations were performed using CPU resources on a workstation equipped with an Intel Core i3-12100 processor (four physical cores and eight hardware threads, operating at a maximum frequency of 4.3~GHz) and 7.5~GiB of physical memory. Offline training-data generation consisted of independent molecular CCSD calculations distributed across two CPU workers to increase throughput. All wall-clock timings reported in the Results and Discussion section correspond to serial runs.

\subsection{Training of the ML module}
The ML training data were generated independently from stretched HF, H$_2$O, N$_2$, and CO geometries. These systems were selected to generate diverse CCSD convergence trajectories at manageable computational cost while exposing the ML model to convergence patterns and amplitude updates across varied electronic environments. Specifically, HF represents a simple yet strongly polar closed-shell diatomic, H$_2$O introduces a bent polyatomic geometry with localized lone pairs and polar bonds, N$_2$ provides a symmetric high-bond-order system that enters increasingly strong-correlation regimes upon stretching, and CO introduces an asymmetric heteronuclear multiple-bond environment with a distinct occupied-virtual orbital structure. Together, these systems span routine closed-shell behavior, polar and asymmetric charge distributions, polyatomic orbital couplings, and strongly stretched regimes, offering a broad set of convergence conditions for training the scalar step-size predictor.

The dataset comprises 70 molecular geometries: 25 HF geometries with H-F distances from 0.80 to 3.20~\AA, 23 symmetric-stretch H$_2$O geometries with O-H distances from 0.80 to 3.00~\AA\ at a fixed H-O-H angle of 104.5$^\circ$, 13 N$_2$ geometries with N-N distances from 0.90 to 3.20~\AA, and 9 CO geometries with C-O distances from 0.95 to 3.00~\AA. All calculations employed the cc-pVDZ basis set with a maximum of 100 CCSD iterations per geometry, yielding 1,532 recorded iterations, of which 76 stall points were identified as valid oracle targets. Thus, supervised targets are concentrated specifically at stall events where recovery is required, while the full trajectories supply the preceding context needed to construct the temporal inputs. Each input sample is built from a sliding window of five consecutive iterations containing the 19 scalar diagnostics defined in Sec.~\ref{feature}, forming a tensor $X_k \in \mathbb{R}^{5 \times 19}$. (Mini-batches up to size 64, i.e., tensors of shape $\le 64 \times 5 \times 19$, are used during training.)

\section{Results and Discussion}

\noindent The four convergence strategies considered in this work are conventional DIIS, pure Newton-Krylov (NK), stall-triggered Newton-Krylov-DIIS (${}{\mathrm{st}}$NK-DIIS), and machine-learning-assisted stall-triggered Newton-Krylov-DIIS (${}{\mathrm{st}}$NK$^{\mathrm{ML}}$-DIIS). The principal computational distinction among the methods is the frequency with which the additional NK is invoked. Pure NK constructs Newton directions throughout the calculation. In contrast, stNK–DIIS invokes this additional work only when sustained DIIS stagnation is detected. The ML-assisted method retains this selective structure and can further reduce the number of fallback trial evaluations when the predicted step is accepted directly. The comparison among all four methods allows us to distinguish three aspects of solver
performance: convergence reliability, recovery efficiency, and computational cost. The methods are evaluated for stretched H$_2$O, CO, CO$_2$, H$_2$CO, CH$_2$NH, BeH$_2$, and N$_2$O geometries using the cc-pVDZ and cc-pVTZ basis sets. The test calculations include molecular systems and geometries that are not identical to those used during training. In particular, the H$_2$O test case has an O-H separation of 3.5 ~\AA\ that lies outside the H$_2$O training range, while CO$_2$, H$_2$CO, CH$_2$NH, BeH$_2$, and N$_2$O represent molecular systems entirely absent from the training set. Additionally, CO at a C-O separation of 4.3 ~\AA\ with cc-pVDZ and at 3.0 ~\AA\ with the larger cc-pVTZ basis set acts as a common molecule probing geometric extrapolation beyond the training range and generalization across basis sets respectively. Collectively, these cases provide an out-of-sample evaluation of the learned step-selection strategy across unseen geometries, molecular systems, and basis sets. \\

\noindent  The stretched configurations are used primarily as challenging numerical test cases for convergence acceleration. Accordingly, the analysis focuses on residual reduction, iteration count, selective ML intervention, and computational efficiency. The resulting CCSD energies are used only as numerical consistency checks and are not interpreted as measures of the physical accuracy of single-reference CCSD for strongly stretched configurations, where enhanced static-correlation or multireference character may become important. The convergence efficiency of the four approaches is examined using both the residual norm as a function of iteration and the residual norm as a function of wall-clock time.\\

\noindent For the wall-clock analysis, the residual histories are mapped onto the measured total wall time of each calculation. For a calculation requiring N iterations with a measured total wall time, T, the time coordinate associated with iteration k is approximated as
\begin{equation}
t_k = \frac{k}{N}T    
\end{equation}
This mapping assumes an approximately uniform distribution of computational cost among iterations and is used only to visualize the convergence trajectories on a common time axis. The reported total wall times are directly measured quantities, whereas the intermediate time coordinates are approximate. This distinction is important because NK iterations generally require substantially more computational work than ordinary DIIS iterations.\\

\noindent We first examine the iteration-wise convergence behavior for stretched H$_2$O with an O-H separation of 3.5 ~\AA\ using the cc-pVDZ basis set. Fig.~1(a) shows the residual norm as a function of iteration number. DIIS initially reduces the residual substantially but subsequently enters a prolonged slow-convergence regime, with the residual remaining predominantly in the $10^{-4}–10^{-5}$ range. After 100 iterations, CCSD is not converged with the residual at approximately $6.6\times10^{-6}$. Pure NK produces a substantially more rapid reduction, reaching the $10^{-7}$ regime in approximately 44 iterations. The stNK-DIIS trajectory initially follows the DIIS behavior and subsequently benefits from selective NK recovery, reaching approximately $9.6\times10^{-8}$ after 90 iterations. The somewhat higher number of iterations appears to be a fallout of returning to the DIIS  procedure which has severe convergence problems. This is observed in other cases as well although ${}{\mathrm{st}}$NK-DIIS manages to converge the equations. The ${}{\mathrm{st}}$NK$^{\mathrm{ML}}$-DIIS calculation follows a similar initial trajectory but exhibits a sharper reduction after entering the difficult convergence regime. The final calculation converges in 46 iterations with a residual of $9.585\times10^{-8}$. The ML prediction does not appear to suffer from remnants of DIIS stagnation in spite of returning to DIIS showing a much sharper convergence trajectory in most cases studied (exception being CO(4.3 ~\AA)/cc-pVDZ where DIIS had stagnated and fluctuated severely at a residual of at $10^{-2}$). The iteration histories therefore demonstrate that selective NK recovery can overcome the prolonged slow-convergence behavior of DIIS, while ML-assisted step selection can further reduce the number of iterations required once the DIIS trajectory becomes ineffective. To complement the iteration-based comparison, Fig~1(b) presents the corresponding residual histories on the approximate wall-time scale. The measured total runtimes are 11.68 s for DIIS (not converged after 100 it.), 33.60 s for pure NK, 15.06 s for ${}{\mathrm{st}}$NK-DIIS, and 4.60 s for ${}{\mathrm{st}}$NK$^{\mathrm{ML}}$-DIIS. Although pure NK converges in substantially fewer iterations, it requires the largest total runtime. The ${}{\mathrm{st}}$NK-DIIS strategy reduces this overhead by restricting NK recovery to the difficult  portion of the trajectory. While the ML-assisted hybrid further reduces the total runtime through cheaper prediction and fewer iterations. For this calculation, the measured runtime of the ML-assisted approach is the lowest. These results illustrate the distinction between iteration efficiency and computational efficiency: fewer iterations do not necessarily imply lower wall-clock cost when each iteration contains expensive Newton-Krylov operations.\\

\noindent Next, we examine CO with a C–O bond length of 4.3~\AA\ using the cc-pVDZ basis set which is particularly challenging for DIIS which terminates after 100 iterations at a residual norm of $4.381\times10^{-2}$, failing to satisfy the prescribed convergence criterion of $10^{-7}$. Fig.~ 1(c) and Fig.~1(d) compare the iteration-wise convergence profiles and approximate wall-clock-time histories of the four methods. The ${\mathrm{st}}$NK$^{\mathrm{ML}}$–DIIS method converges in 97 iterations and 36.09 s, compared with 80 iterations and 40.73 s for ${\mathrm{st}}$NK–DIIS and 56 iterations and 104.67 s for pure NK. Although ${\mathrm{st}}$NK$^{\mathrm{ML}}$–DIIS requires more iterations than either pure NK or ${\mathrm{st}}$NK–DIIS, its substantially lower wall-clock time demonstrates its efficiency. The shorter runtime relative to ${\mathrm{st}}$NK–DIIS also suggests that the ML-predicted correction magnitude can improve the efficiency of the NK recovery process. To demonstrate transferability across basis sets for the ML- variant, we next study Fig.~1(e) and Fig.~1(f) which show results for CO (C-O = 3.0~\AA{}/cc-pVTZ) while the training set included CO (C-O = 3.0~~\AA\/cc-pVDZ). 
The ${\mathrm{st}}$NK$^{\mathrm{ML}}$–DIIS method converges easily here.
This was expected as our ML model requires no electronic structure information, relying only on convergence pattern detection.
All four methods converge, with the ${\mathrm{st}}$NK$^{\mathrm{ML}}$–DIIS needing 438.42 s, compared with 644.71 s for DIIS, 1292.25 s for pure NK, and 1080.48 s for ${\mathrm{st}}$NK–DIIS. Unlike the preceding calculations, conventional DIIS converges successfully in this case, even so, its convergence trajectory exhibits stagnation, triggering the stall-detection criterion and activating NK correction calculations in both hybrid methods. The advantage of ${\mathrm{st}}$NK$^{\mathrm{ML}}$–DIIS in this case therefore lies not in recovering convergence from DIIS failure, but in improving computational efficiency when conventional DIIS already converges. Compared with DIIS, $\mathrm{st}$NK-DIIS shows improved stability. The DIIS residual stalls twice. Once quite early in the run near $1.7\times10^{-4}$, where the NK step lowers it to $1.1\times10^{-4}$, and again near $9\times10^{-6}$ for about five iterations ($9.49\times10^{-6}$ to $9.14\times10^{-6}$). The residual of $\mathrm{st}$NK-DIIS decreases steadily through this region. The method therefore breaks stagnation, but the reduction obtained at the stall is modest, and the second stall still occurs, so that in the final stages $\mathrm{st}$NK-DIIS converges more slowly than DIIS. However, the ML-assisted step selection, is able to avoid the second stall completely with only one ML extrapolation. The iteration counts provide further insight into this improvement. DIIS requires 64 iterations (644.71s), pure NK requires 19 (1292.25s), ${\mathrm{st}}$NK–DIIS requires 68 (1080.48s), and ${\mathrm{st}}$NK$^{\mathrm{ML}}$–DIIS requires 44 (438.42s). This comparison provides evidence that the proposed ML-assisted hybrid can improve computational efficiency even when conventional DIIS converges successfully. \\

\noindent Fig.~2(a) shows the iteration-dependent convergence behavior of CO$_2$ at a C-O separation of 2.4 ~\AA\ using the cc-pVDZ basis set. This system is not included in the ML training set. Although it shares the C-O bonding motif with the trained CO system and some qualitative near-degeneracy characteristics with stretched H$_2$O, CO$_2$ has a distinct molecular geometry and electronic structure. Its stretched geometry therefore provides a useful cross-system test of ML step-size transferability. DIIS achieves a substantial initial reduction but subsequently develops an extended slow-convergence regime, with the residual remaining near the $10^{-3}$ level before gradually decreasing toward the $10^{-5}$ range. After 100 iterations, the residual remains approximately $3.999\times10^{-6}$ (419.19s). Pure NK produces a more rapid and nearly monotonic reduction, reaching approximately $6.469\times10^{-8}$ within 30 iterations (829.58s). The ${}{\mathrm{st}}$NK-DIIS approach initially follows the DIIS trajectory and subsequently introduces NK recovery as convergence becomes limited, converging after 82 iterations (652.03s). The ${}{\mathrm{st}}$NK$^{\mathrm{ML}}$-DIIS trajectory  exhibits a sharper reduction in the difficult convergence regime beginning near iteration number 44. It reaches approximately $9.811\times10^{-8}$ after 61 iterations (273.54s). Fig. 2(b) shows the corresponding residual histories on the approximate wall-time scale following the order pure NK $>$ ${}{\mathrm{st}}$NK-DIIS $>$ DIIS $>$ ${}{\mathrm{st}}$NK$^{\mathrm{ML}}$-DIIS. 

\noindent We also tested H$_2$CO with a C-O separation of 3.0~~\AA\ and CH$_2$NH with a C-N separation of 3.2~~\AA\, both using the cc-pVDZ basis set. The results, provided in the Supporting Information, show similar behaviour as the cases discussed above.\\

\noindent Next, we consider linear BeH$_2$ with a Be–H bond length of 5.5~\AA{} using the cc-pVTZ basis set. Neither this molecular system nor the basis set was included in the ML training data, providing a test of generalization to an unseen molecular system and basis set. Fig.~2(c) shows that the DIIS trajectory exhibits substantial initial fluctuations, with the residual norm reaching $2.323\times10^{4}$ before decreasing progressively and eventually converging after 52 iterations. In contrast, pure NK reduces the residual rapidly and achieves convergence in only six iterations. The ${\mathrm{st}}$NK–DIIS method reaches a residual norm below $10^{-7}$ after 50 iterations, whereas ${\mathrm{st}}$NK$^{\mathrm{ML}}$–DIIS follows a similar initial trajectory but subsequently reduces the residual more rapidly, converging in 42 iterations with a final residual norm of $8.412\times10^{-8}$. These results highlight the efficiency of the ML-assisted hybrid over both DIIS and the non-ML hybrid for a molecule and basis set excluded from training set. To assess computational efficiency, Fig. 2(d) presents the measured wall-clock times for the four methods: 262.74 s for DIIS, 187.31 s for pure NK, 315.23 s for ${\mathrm{st}}$NK–DIIS, and 187.31 s for ${\mathrm{st}}$NK$^{\mathrm{ML}}$–DIIS. The ML-assisted hybrid therefore reduces the wall-clock time by approximately 28.7$\%$ relative even to DIIS. Although ${\mathrm{st}}$NK$^{\mathrm{ML}}$–DIIS does not outperform pure NK in wall-clock time for this test, it achieves comparable computational efficiency. The residual histories also demonstrate the robustness of the stall-triggered strategy. The DIIS residual rises to $2.323\times10^{4}$ at the second iteration and remains as high as $2.5\times10^{3}$ afterwards. The residual of $\mathrm{st}$NK-DIIS consistently remains more than three orders of magnitude below the DIIS peak. 
Although DIIS eventually converges on its own, in this case, $\mathrm{st}$NK-DIIS can act as a safeguard, suppressing large residual excursions at modest additional cost. In other cases, DIIS may not be able to recover. Incorporating the ML-predicted step length retains this robustness while making the method faster than both DIIS and $\mathrm{st}$NK-DIIS.\\


\begin{table}[H] 
\centering
\caption{Summary of DIIS, pure NK, ${}{\mathrm{st}}$NK-DIIS, and ${}{\mathrm{st}}$NK$^{\mathrm{ML}}$-DIIS calculations.} 
\label{tab:results} 
\resizebox{\textwidth}{!}{%
\begin{tabular}{llc cccc c cccc}
\toprule
& & & \multicolumn{4}{c}{No. of Iterations}
& & \multicolumn{4}{c}{Wall Time (s)} \\
\cmidrule(lr){4-7}
\cmidrule(lr){9-12}
System & Geometry (~\AA) & Basis set
& DIIS & NK & stNK-DIIS & stNK$^{\mathrm{ML}}$-DIIS
& ML Steps
& DIIS & NK & stNK-DIIS & stNK$^{\mathrm{ML}}$-DIIS \\
\midrule
H$_2$O & O-H = 3.5 & cc-pVDZ& 100 & 44 & 90 & 46 & 1&11.68& 33.60 & 15.06 & 4.60 \\
& & & ($6.6\times10^{-6}$) & & & & & & & &  \\
CO & C-O = 4.3& cc-pVDZ&100&56&80&97&2 & 27.29 & 104.67& 40.73 & 36.09 \\
& & & ($4.38\times10^{-2}$) & & & & & & & &  \\
CO & C-O = 3.0& cc-pVTZ&64 & 19 & 68 & 44 & 1 &644.71 & 1292.25 & 1080.48$^{a}$&438.42\\
& & & ($8.7\times10^{-8}$) & & & & & & & &  \\
CO$_2$ & C-O = 2.4& cc-pVDZ& 100 &30 & 82 &61&2&419.19& 829.54& 652.03&273.58 \\ 
& & & ($3.99\times10^{-6}$) & & & & & & & &  \\
H$_2$CO$^{b}$ & C-O = 3.0& cc-pVDZ&100 & 54 & 77 &45 & 1 &196.87 & 898.49&187.71& 78.04  \\ 
& & & ($2.16\times10^{-7}$) & & & & & & & &  \\
CH$_2$NH$^{b}$ & C-N = 3.2& cc-pVDZ&100 & 23 & 45 & 41 & 1 & 354.66&624.33 &256.88&146.26 \\
& & & ($1.55\times10^{-7}$) & & & & & & & &  \\
BeH$_2$  & Be-H = 5.5 & cc-pVTZ &52 & 06 & 50 &42 & 1 & 262.74& 187.31& 315.23$^{c}$& 187.31\\
& & & ($5.47\times10^{-8}$) & & & & & & & &  \\
N$_2$O   & N-O = 2.5 & cc-pVDZ &100 & 22 & 109$^{d}$ &84 & 3&238.23 &360.21&426.28&243.79  \\ 
&\& N-N=2.5 & &($5.86\times10^{-4}$) & & & &  \\ 
\bottomrule 
\end{tabular} 
}
\vspace{2mm}
\begin{minipage}{0.9\textwidth}
\footnotesize
$^{a}$ Improved residual stability at additional computational cost; motivates the ML variant.\\
$^{b}$ Detailed results are provided in supplementary information.\\
$^{c}$ Stabilization of residual blow-up comes at the cost of increased wall time.\\
$^{d}$ Demonstrates convergence recovery within the DIIS framework. 
\end{minipage}
\end{table}

\noindent For N$_2$O, we consider a linear geometry with N–N and N–O bond lengths of 2.5~~\AA\ and a bond angle of $180^\circ$, using the cc-pVDZ basis set. As shown in Fig.~2(e), conventional DIIS stagnates at a residual norm of $5.863\times10^{-4}$, failing to meet the convergence criterion. In contrast, pure NK converges within 22 iterations, while the ${\mathrm{st}}$NK–DIIS approach also achieves convergence but requires 109 iterations. Notably, incorporating ML-guided step-size selection substantially reduces the iteration count to 84 relative to the non-ML hybrid, highlighting its potential to accelerate convergence while retaining the robustness of the underlying framework. Moreover, as shown in Fig.~2(f), the ${\mathrm{st}}$NK$^{\mathrm{ML}}$–DIIS method converges in 243.79 s, compared with 360.21 s for pure NK, yielding a 32.3$\%$ reduction in wall-clock time. In contrast, ${\mathrm{st}}$NK–DIIS requires 426.28 s, exceeding the runtime of pure NK. Nevertheless, it successfully recovers convergence in a case where DIIS fails. The ML-guided step-size selection is thus significant for maintaining the efficiency of this NK-guided recovery strategy. These results highlight the potential of ML-guided Newton-step magnitude selection to accelerate convergence while preserving the robustness of the residual-verified NK-DIIS hybrid framework.
\begin{figure}[H]
\centering

\begin{subfigure}{0.48\textwidth}
    \centering
    \includegraphics[width=\textwidth]{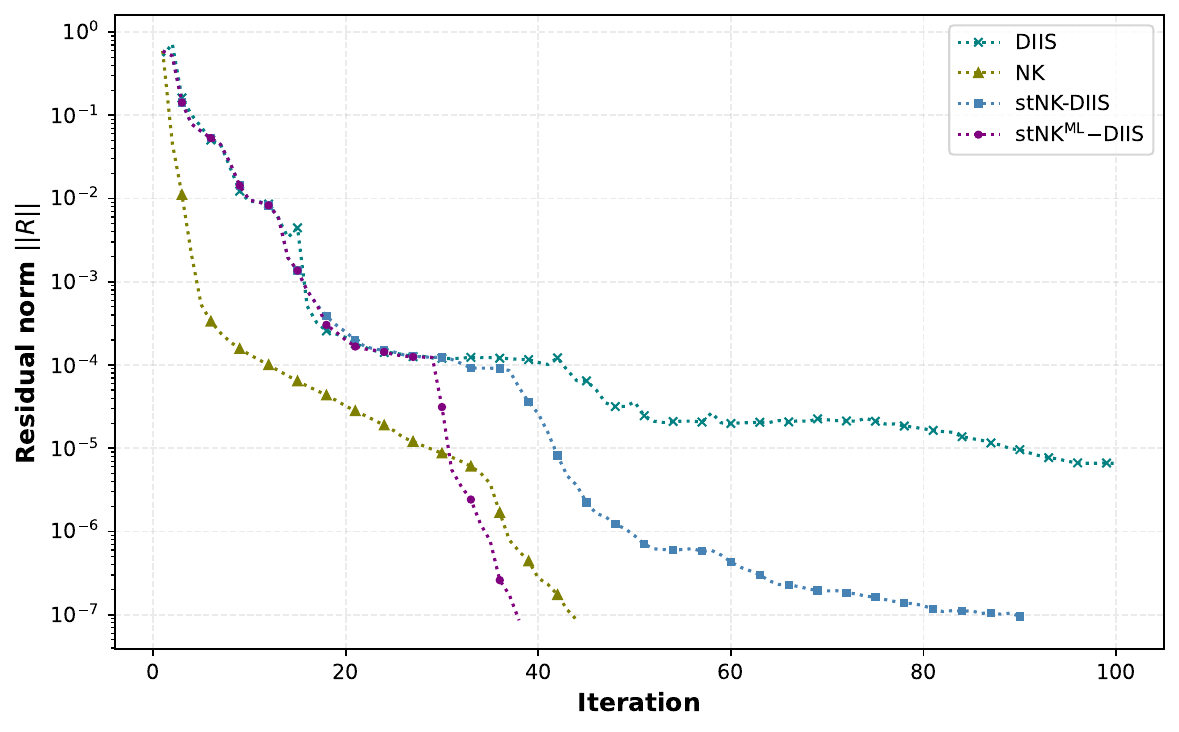}
    \caption{H$_2$O: Iteration.}
    \label{fig:h2o_iteration}
\end{subfigure}
\begin{subfigure}{0.48\textwidth}
    \centering
    \includegraphics[width=\textwidth]{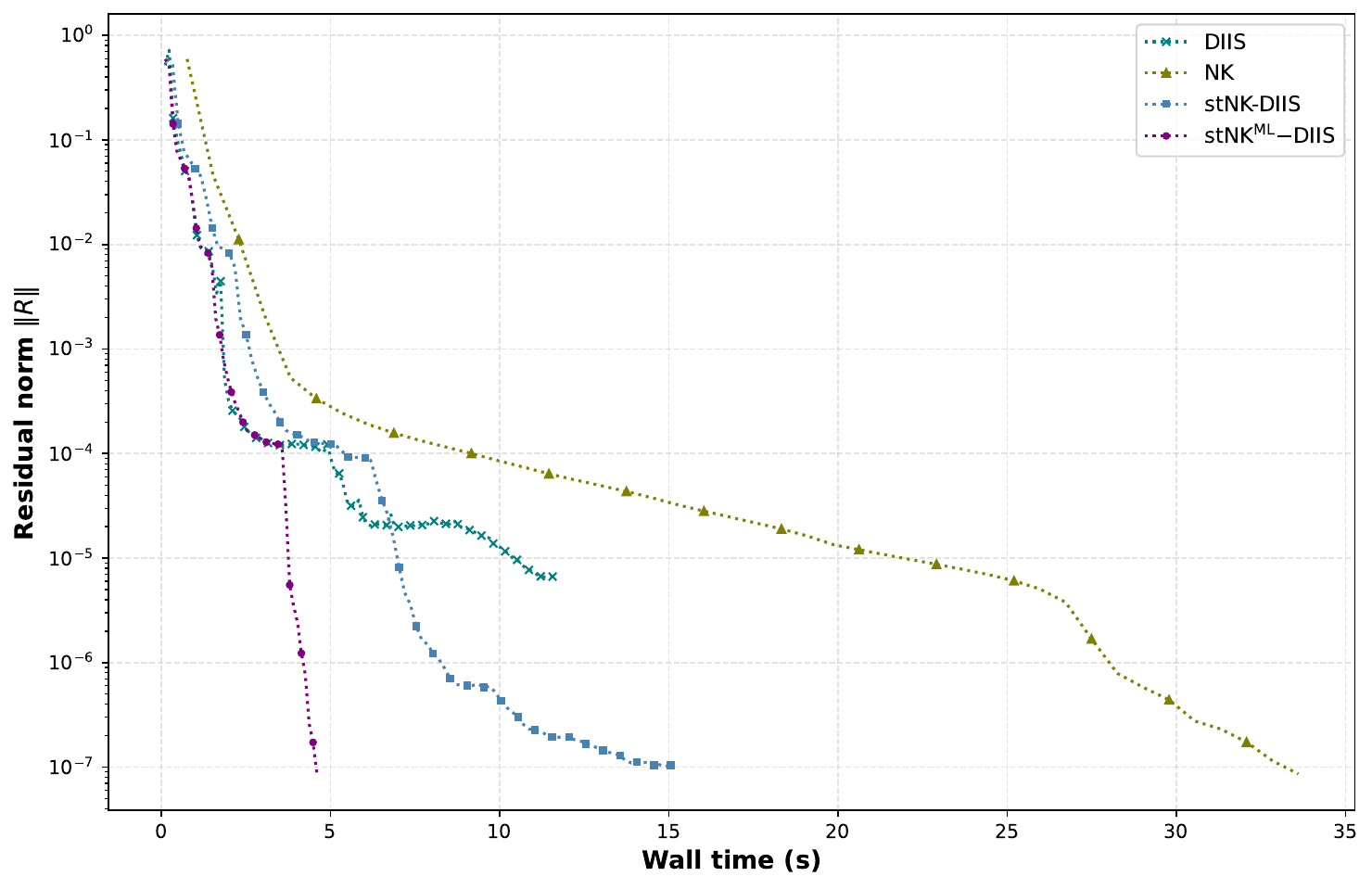}
    \caption{H$_2$O: Wall time.}
    \label{fig:h2o_time}
\end{subfigure}

\vspace{0.1cm}

\begin{subfigure}{0.48\textwidth}
    \centering
    \includegraphics[width=\textwidth]{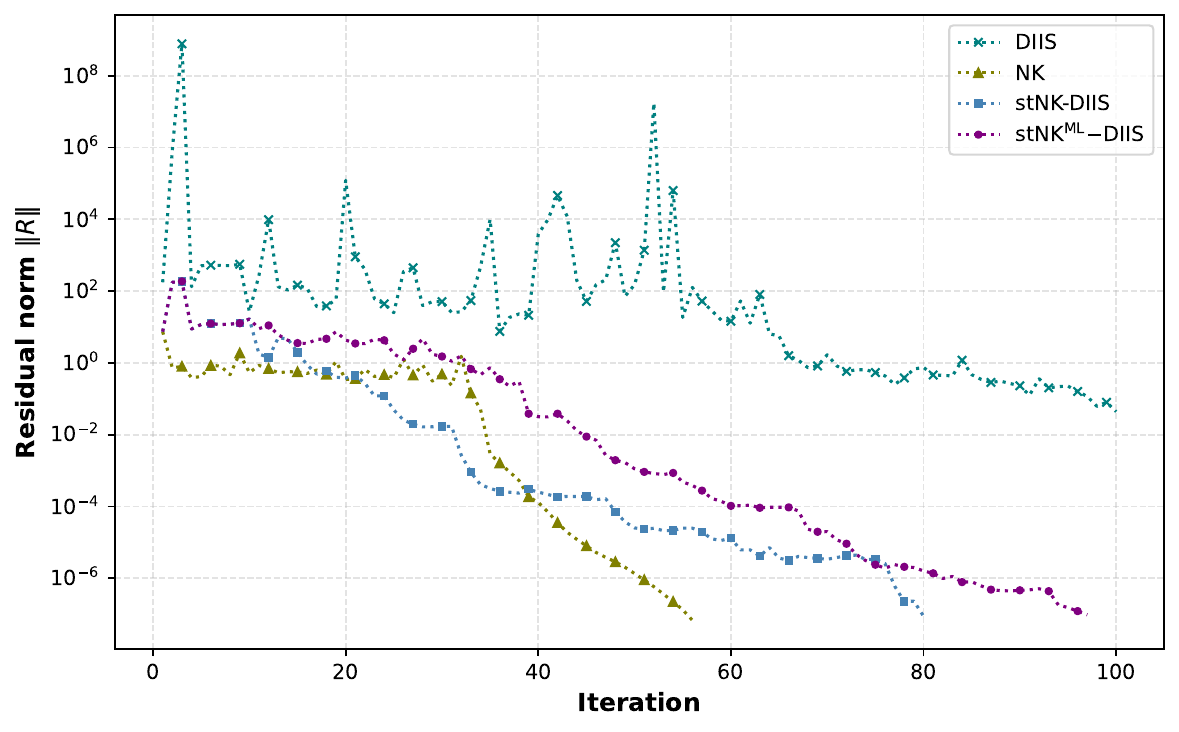}
    \caption{CO (cc-pVDZ): Iteration.}
    \label{fig:co_iteration}
\end{subfigure}
\begin{subfigure}{0.48\textwidth}
    \centering
    \includegraphics[width=\textwidth]{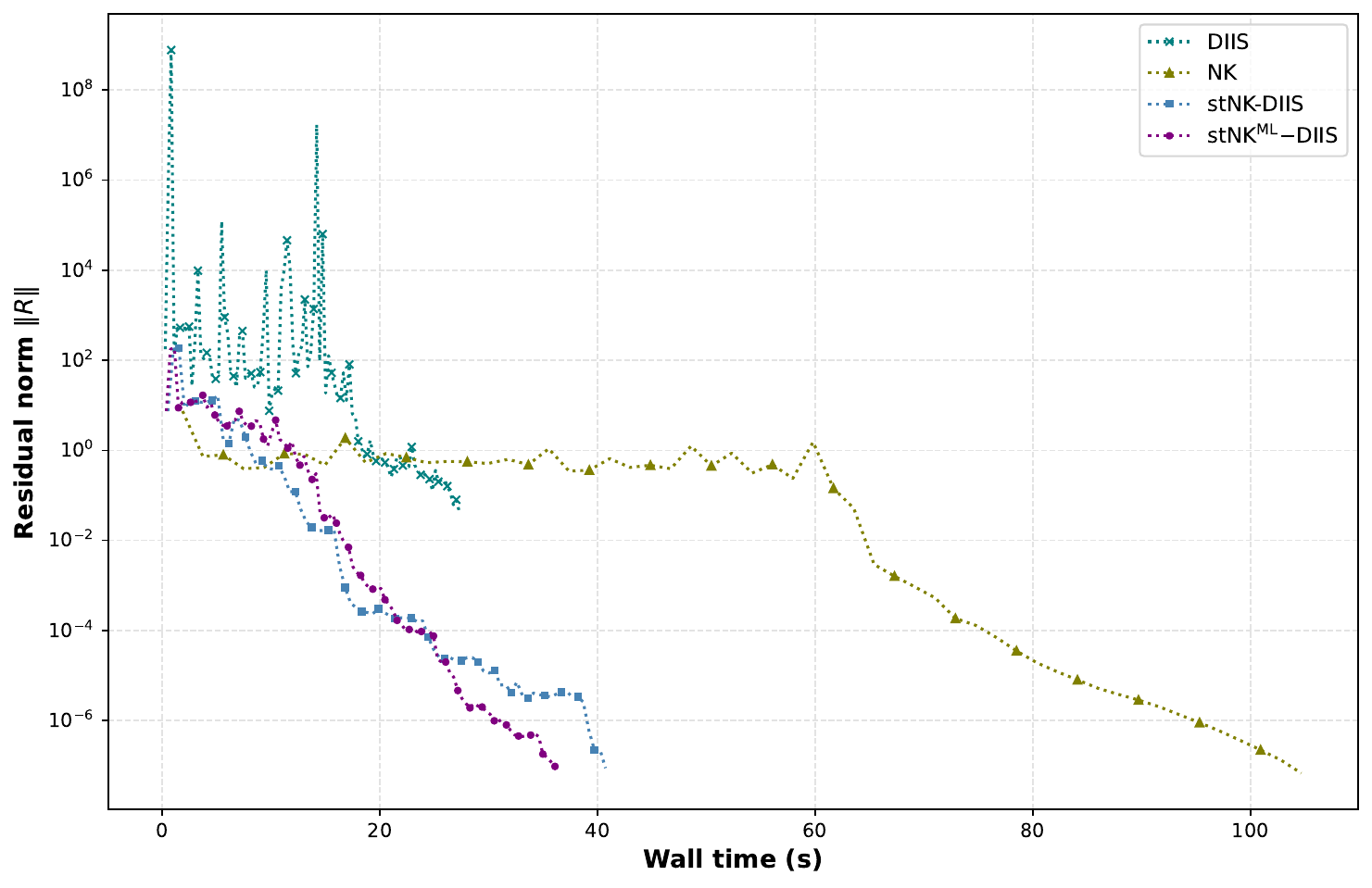}
    \caption{CO (cc-pVDZ):  Wall time.}
    \label{fig:co_walltime}
\end{subfigure}

\vspace{0.1cm}

\begin{subfigure}{0.48\textwidth}
    \centering
    \includegraphics[width=\textwidth]{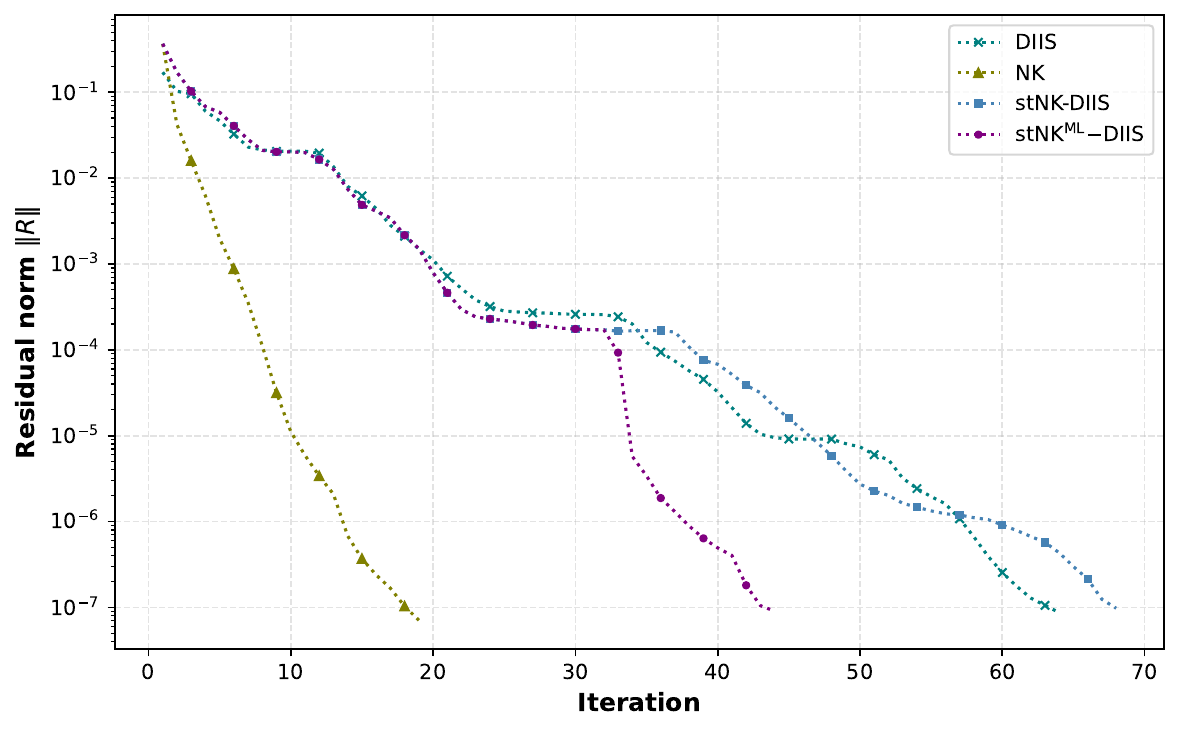}
    \caption{CO (cc-pVTZ): Iteration.}
    \label{fig:co_pvtz_iteration}
\end{subfigure}
\begin{subfigure}{0.48\textwidth}
    \centering
    \includegraphics[width=\textwidth]{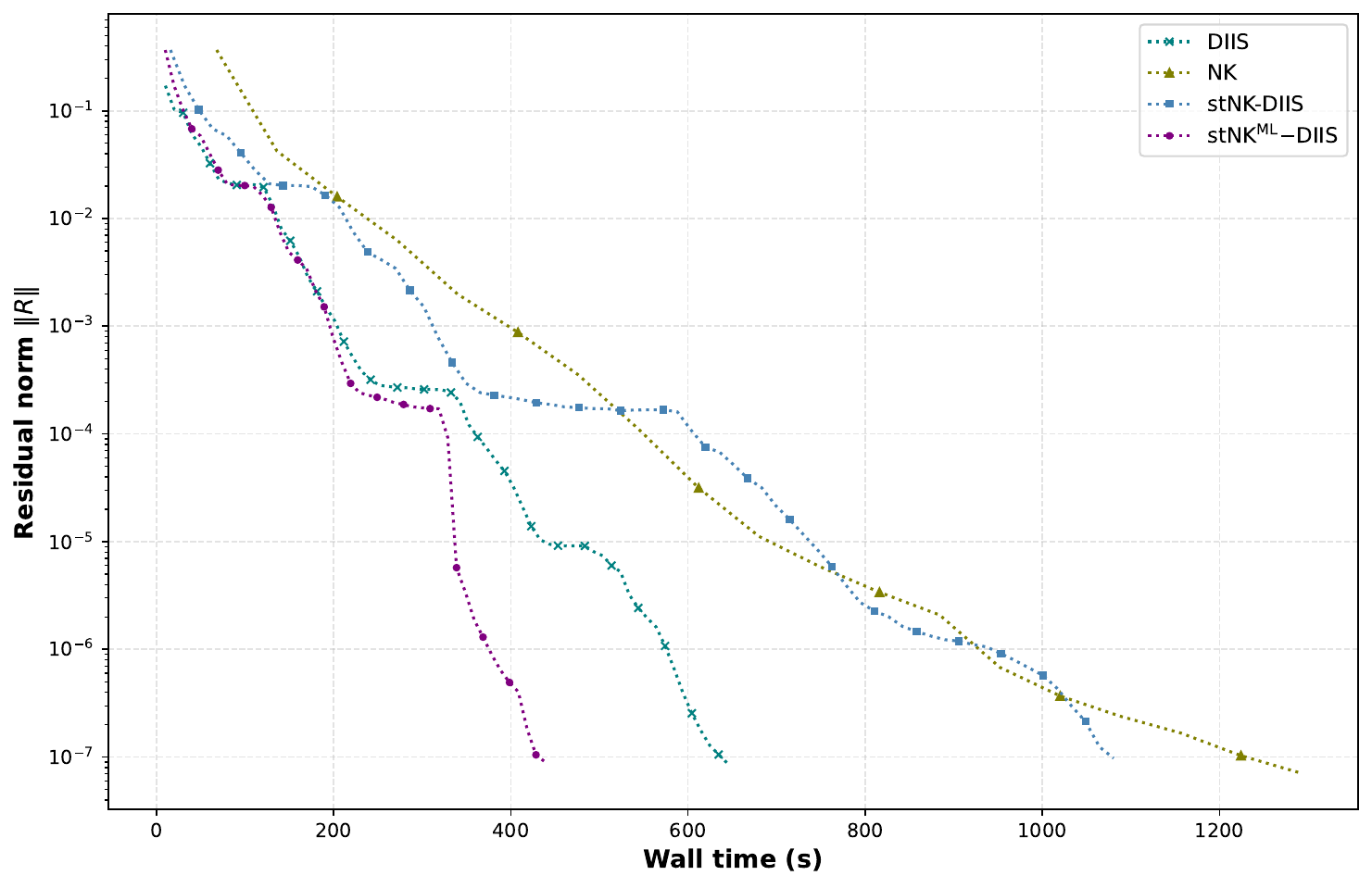}
    \caption{CO (cc-pVTZ):  Wall time.}
    \label{fig:co_pvtz_walltime}
\end{subfigure}

\vspace{0.1cm}

\caption{Comparison of CCSD residual-norm convergence for H$_2$O and CO using DIIS, pure NK, stNK-DIIS, and ${\mathrm{st}}$NK$^{\mathrm{ML}}$-DIIS. The left column shows residual-norm decay versus iteration, while the right column shows residual-norm decay versus approximate wall time.}
\label{fig:all_molecular_convergence}

\end{figure}

\begin{figure}[H]
\centering

\begin{subfigure}{0.48\textwidth}
    \centering
    \includegraphics[width=\textwidth]{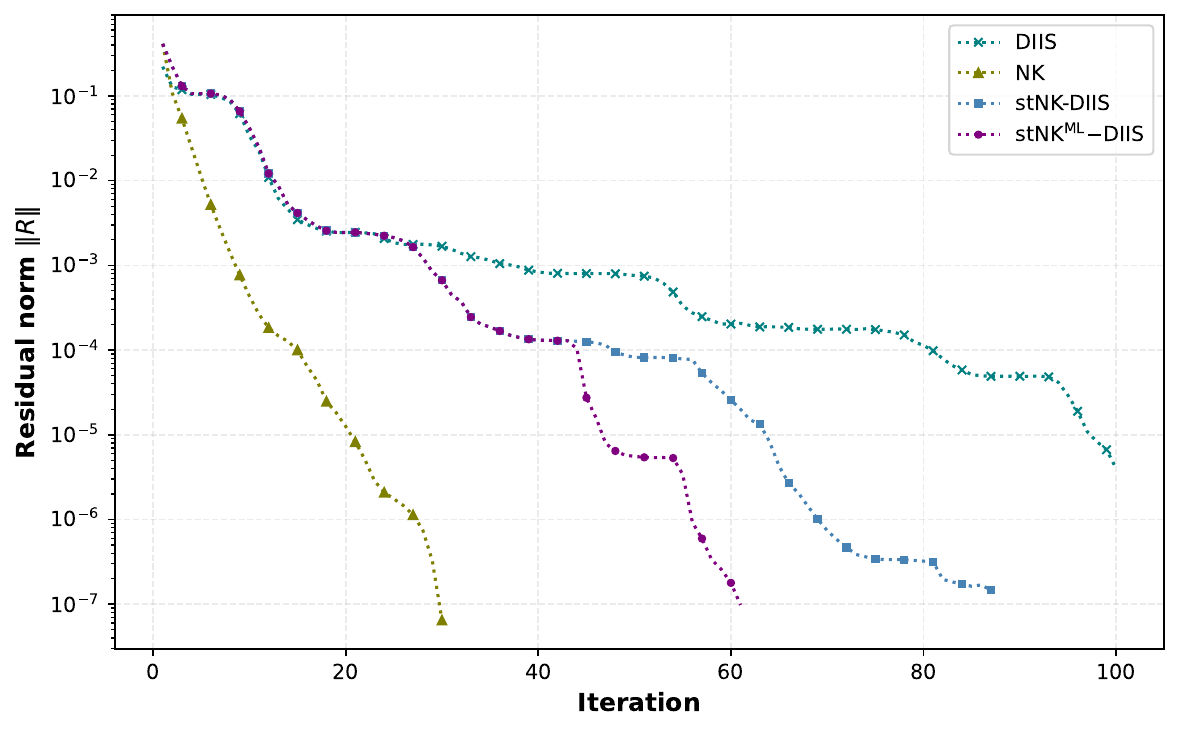}
    \caption{CO$_2$: Iteration.}
    \label{fig:co2_iteration}
\end{subfigure}
\begin{subfigure}{0.48\textwidth}
    \centering
    \includegraphics[width=\textwidth]{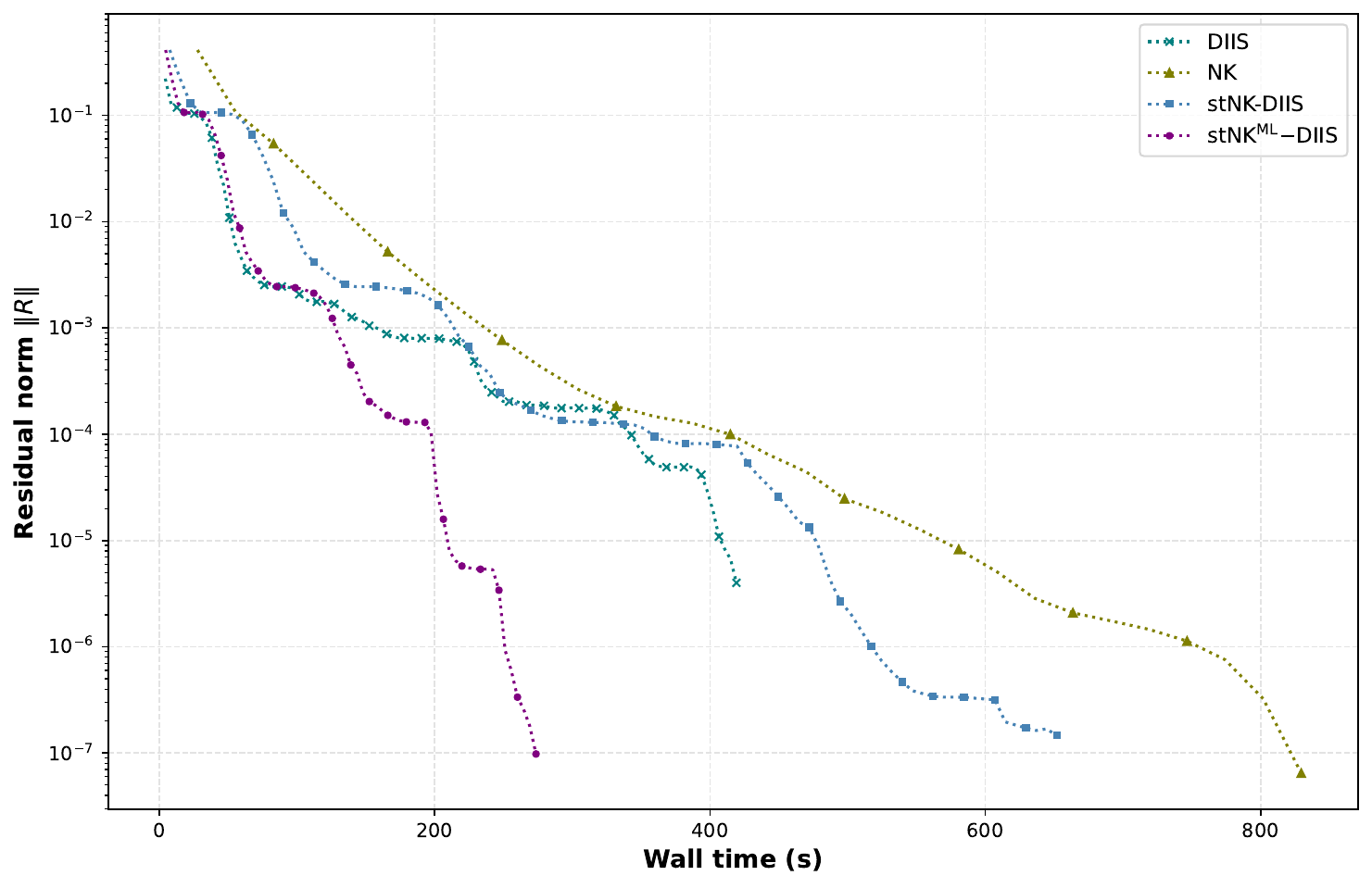}
    \caption{CO$_2$:  Wall time.}
    \label{fig:co2_walltime}
\end{subfigure}

\vspace{0.1cm}

\begin{subfigure}{0.48\textwidth}
    \centering
    \includegraphics[width=\textwidth]{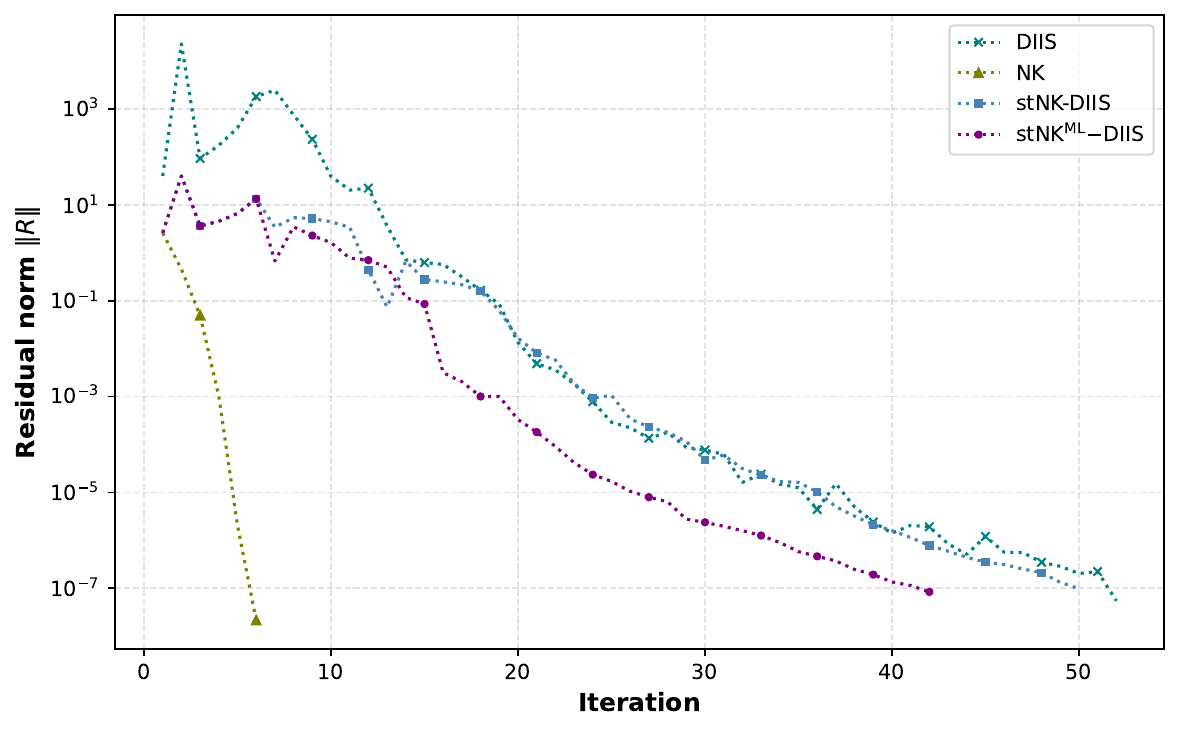}
    \caption{BeH$_2$: Iteration.}
    \label{fig:beh2_iteration}
\end{subfigure}
\begin{subfigure}{0.48\textwidth}
    \centering
    \includegraphics[width=\textwidth]{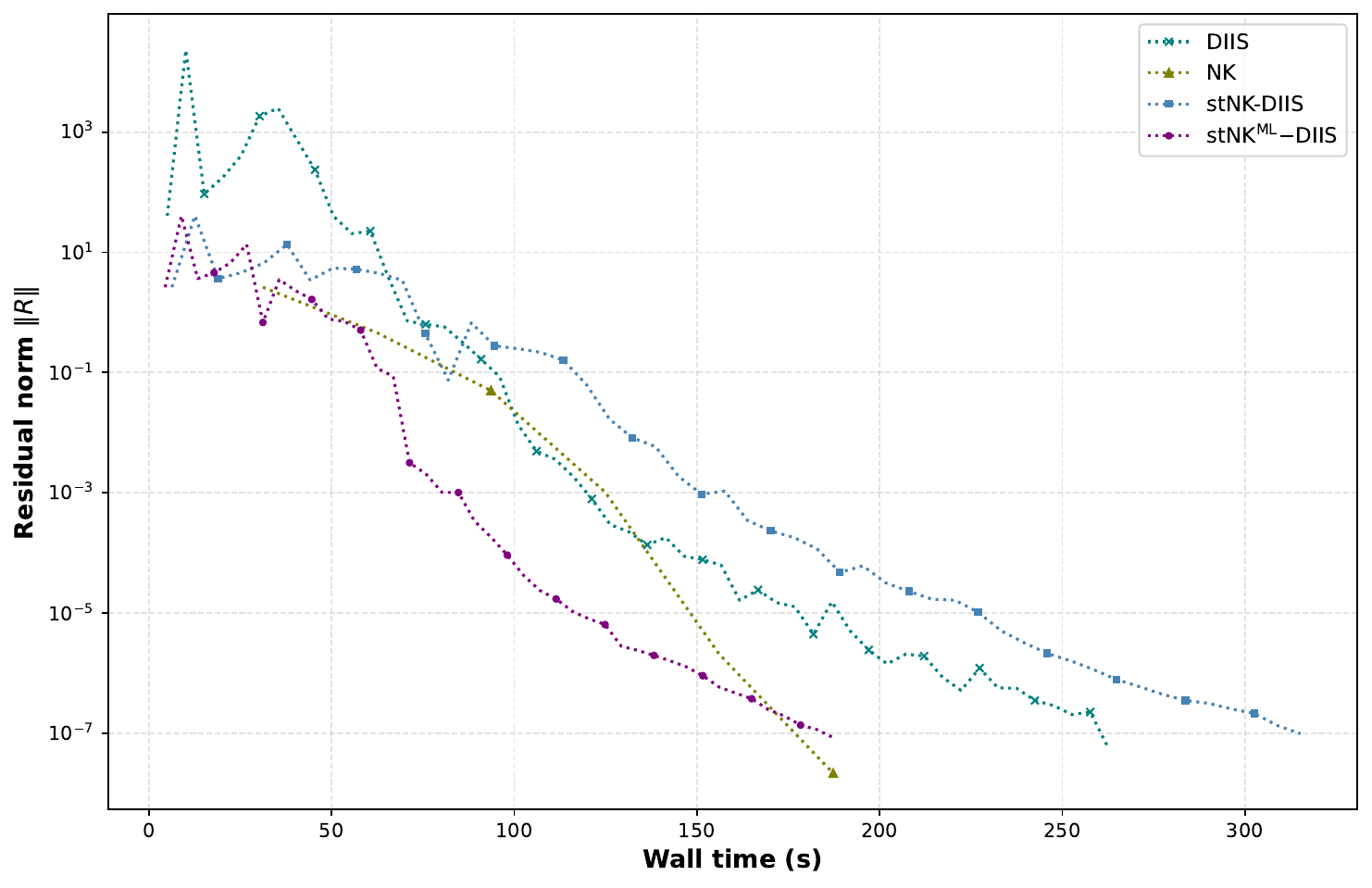}
    \caption{BeH$_2$:  Wall time.}
    \label{fig:beh2_walltime}
\end{subfigure}

\vspace{0.1cm}

\begin{subfigure}{0.48\textwidth}
    \centering
    \includegraphics[width=\textwidth]{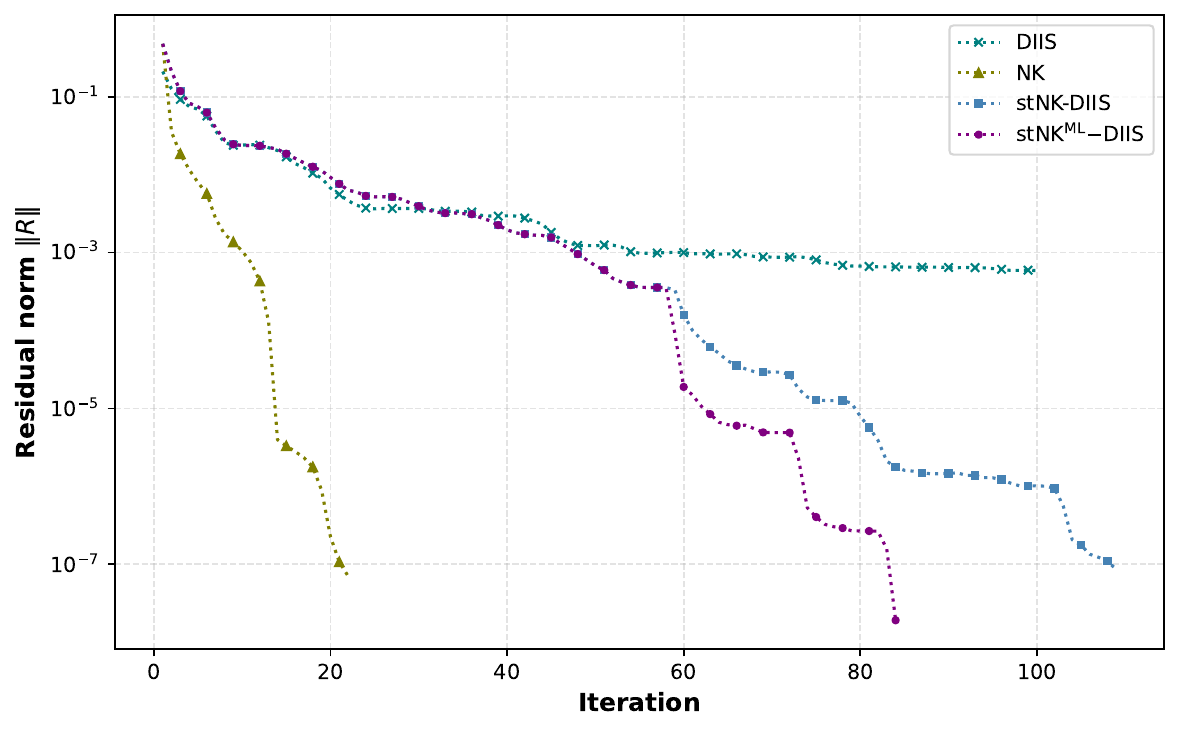}
    \caption{N$_2$O: Iteration.}
    \label{fig:n2o_iteration}
\end{subfigure}
\begin{subfigure}{0.48\textwidth}
    \centering
    \includegraphics[width=\textwidth]{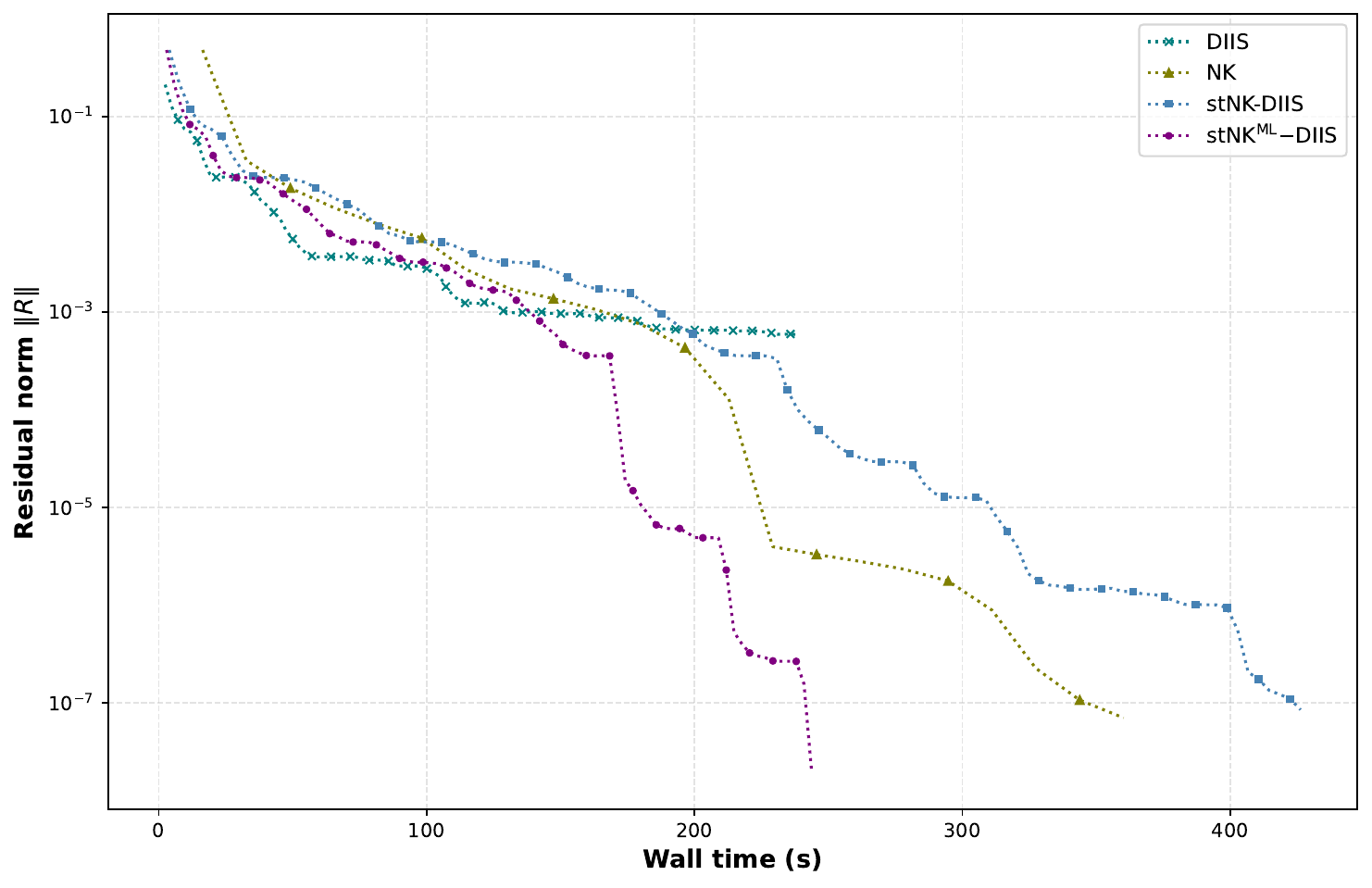}
    \caption{N$_2$O:  Wall time.}
    \label{fig:n2o_walltime}
\end{subfigure}

\caption{Comparison of CCSD residual-norm convergence for CO$_2$, BeH$_2$, N$_2$O using DIIS, pure NK, stNK-DIIS, and ${\mathrm{st}}$NK$^{\mathrm{ML}}$-DIIS. The left column shows residual-norm decay versus iteration, while the right column shows residual-norm decay versus approximate wall time.}
\label{fig:all_molecular_convergence}

\end{figure}

\noindent The four convergence strategies, thus, exhibit distinct trade-offs in iteration count, per-iteration computational cost, and total wall-clock time. To summarize, conventional DIIS is computationally inexpensive during routine iterations and converges successfully for CO at a C–O bond length of 3.0~\AA{} with cc-pVTZ and BeH$_2$ with cc-pVTZ. However, it exhibits prolonged stagnation for stretched H$_2$O and CO$_2$, and fails to satisfy the convergence criterion for stretched CO at 4.3~\AA{} and N$_2$O. Pure NK generally reaches the convergence threshold in fewer iterations, particularly for BeH$_2$, N$_2$O, and CO with cc-pVTZ although at higher computational cost per iteration. Consequently, despite its lower iteration count, pure NK incurs longer runtimes unless the number of iterations is extremely low. The $\mathrm{st}$NK-DIIS method combines the low cost of DIIS with selective NK recovery and enables convergence in all cases where DIIS stagnates or fails, as demonstrated for stretched H$_2$O, CO, CO$_2$, and N$_2$O. However, in cases where DIIS converges even with some intermediate difficulty, its wall time can exceed that of pure DIIS (CO and BeH$_2$ in cc-pVTZ) although the convergence trajectory becomes more stable. Specifically, DIIS exhibits a large residual blow-up for BeH$_2$ and slow progress through plateaus for CO, whereas $\mathrm{st}$NK-DIIS suppresses the former and mitigates the latter at the cost of higher computational cost. This observation motivates the incorporation of ML-assisted step-size selection to improve the efficiency of the NK recovery. Indeed, $\mathrm{st}$NK$^{\mathrm{ML}}$-DIIS achieves the shortest wall time (42 iterations) among the four methods for almost all of the reported systems, including stretched H$_2$O, CO, CO$_2$, and N$_2$O. The only exception is BeH$_2$, where it matches the wall time of pure NK (which needs only 6 iterations) while outperforming DIIS and $\mathrm{st}$NK-DIIS. Additional results for H$_2$CO and CH$_2$NH, presented in the Supporting Information, further demonstrate the applicability of the approach across different molecular systems and geometries.\\

\noindent Table~1 summarizes the performance of ${\mathrm{st}}$NK$^{\mathrm{ML}}$-DIIS across the tested systems. In every case, the method converges to a CCSD residual norm below $1.0\times10^{-7}$, requiring only one to three ML-assisted step selections per calculation. This limited number of interventions indicates that the learned component is used selectively to guide the correction magnitude when the DIIS trajectory enters a stalled regime, rather than replacing the underlying iterative solver or altering the CCSD equations. 
The proposed method appears to offer a favorable balance between convergence robustness and computational cost across the tested systems and basis sets. Although its performance advantage depends on the system and it is not universally faster than pure NK (if NK requires very few iterations), the results support the use of ML-guided step-size selection as a promising strategy for accelerating challenging CCSD convergence behaviour.

\section{Conclusions}
A hybrid convergence-acceleration framework for CCSD was developed and implemented in an original, standalone Python solver suite interfaced with Psi4. The approach combines conventional DIIS with selective Newton-Krylov (NK) recovery and, in its ML-assisted formulation, employs a GRU model to predict the scalar step magnitude along the computed Newton direction. This design assigns distinct roles to each component: the Jacobi update provides the diagonal baseline, DIIS accelerates convergence using iteration history, NK supplies a locally informed recovery direction when DIIS stagnates, and the ML model guides the magnitude of the resulting correction. Because training relies solely on five-iteration temporal windows of 19 scalar convergence diagnostics with oracle step-size targets evaluated only at detected DIIS stalls, the model incurs minimal training and storage overhead, and full NK trajectories are not required even during data generation.

Calculations for stretched H$_2$O, CO, CO$_2$, H$_2$CO, CH$_2$NH, BeH$_2$, and N$_2$O demonstrate the effectiveness and robustness of the proposed strategies. The $\mathrm{st}\text{NK-DIIS}$ approach overcomes DIIS stagnation and divergence while limiting the computational overhead of pure NK by invoking Krylov solves selectively, although its total runtime remains system-dependent. The ML-assisted variant ($\mathrm{st}\text{NK}^{\mathrm{ML}}\text{-DIIS}$) further improves recovery efficiency in most tested cases, achieving residual norms below $10^{-7}$ with only one to three ML-assisted step selections per calculation and delivering the best overall wall-clock performance. These results underscore the importance of evaluating wall-clock time alongside iteration count when assessing convergence-acceleration schemes. More broadly, they demonstrate how machine learning can strengthen a numerical solver without altering its underlying equations. 

\section*{Data Availability Statement}

The source codes, input files, pre-trained model checkpoint, and
training trajectory data supporting the findings of this study are
openly available in the Zenodo repository at
\href{https://doi.org/10.5281/zenodo.24376963}{DOI:~10.5281/zenodo.24376963}
(version~v1.0.0, \url{https://zenodo.org/records/24376963}) and on
GitHub at \url{https://github.com/shamikiiser/stNK-DIIS-CCSD}.
The repository contains the implementations of conventional DIIS,
pure Newton-Krylov (NK), stall-triggered Newton-Krylov-assisted DIIS
($\mathrm{st}\text{NK-DIIS}$), and its machine-learning-assisted variant
($\mathrm{st}\text{NK}^{\mathrm{ML}}\text{-DIIS}$), together with the data-harvesting and GRU training
scripts, the pre-trained step-ratio predictor (\texttt{cc\_accelerator.pt}),
scalar convergence trajectory datasets (\texttt{cc\_training\_data/}),
and example molecular input files required to reproduce the
calculations reported in this work.

\section*{Supporting Information}
Additional convergence results for the H$_2$CO and CH$_2$NH systems, along with supplementary figures and computational details, are available in the Supporting Information.

\begin{acknowledgement}
The authors express their gratitude to IISER Kolkata for providing the necessary research infrastructure. S.B. gratefully acknowledges funding support from IISER Kolkata, and S.C. is grateful to IISER Kolkata for the Senior Research Fellowship. The authors also acknowledge the National Supercomputing Mission (NSM) for providing additional computational resources through PARAM RUDRA, hosted at the S. N. Bose National Centre for Basic Sciences, implemented by C-DAC, and supported by the Ministry of Electronics and Information Technology (MeitY) and the Department of Science and Technology (DST), Government of India.
\end{acknowledgement}

\section*{Author Declarations}
\subsection*{Conflict of Interest}
The authors have no conflicts to disclose.
\subsection*{Author Contributions}
S.C. and S.B. contributed equally to this work.

\bibliography{references}

\end{document}